\def\pr #1 #2 #3 {{\sl Phys. Rev.} {\bf #1}, #2 (#3)}
\def\prl #1 #2 #3 {{\sl Phys. Rev. Lett.} {\bf #1}, #2 (#3)}
\def\zp #1 #2 #3 {{\sl Z. Phys.} {\bf #1}, #2 (#3)}

\documentstyle[prl,aps,multicol,epsfig]{revtex}
\begin{document}
\draft
\title{\Large Optical response in one dimensional Mott Insulators}
\author{S. S. Kancharla and C. J. Bolech}
\address{Serin Physics Laboratory, Rutgers University, Piscataway, New
Jersey 08854-8019, USA}
\date{December 22, 2000}

\maketitle

\begin{abstract}
We study the optical response of a Mott Hubbard system in the
framework of the half--filled Extended Hubbard Model using the Density Matrix
Renormalization Group (DMRG) method. We discuss the appearance of
excitonic features inside the spectral gap as the system goes from the
Spin Density Wave (SDW) to the Charge Density Wave (CDW) phase.
\end{abstract}

\begin{multicols}{2}
\narrowtext

A detailed understanding of the optical response and charge gap in the
one dimensional Mott insulators remains a challenge to existing theoretical
methods. Renewed interest in the subject stems from several
experiments on materials such as SrCuO$_2$ and Ni halides\cite{mizuno,kishida} with
possible applications such as ultrafast switching in optoelectronic
devices. The aim of this letter is to investigate the effect of
nearest neighbor Coulomb repulsion on the optical and Raman spectrum
in these systems.

A simplified model which has been used to describe the essential
physics of these materials is the Extended Hubbard Model (EHM)
defined as,
\begin{eqnarray}
H&=&-t\sum_{j,\sigma}\left( c_{j+1\sigma}^{\dagger }c_{j\sigma}+ 
                                     {\em h.c.} \right) \nonumber \\ 
 && + U\sum_{j}\left(n_{j\uparrow}   - \frac{1}{2}\right)
               \left(n_{j\downarrow} - \frac{1}{2}\right) \nonumber \\ 
 && + V\sum_{j}\left(n_{j}-1\right) \left(n_{j+1}-1\right)
\label{exthubbard}
\end{eqnarray} 

The first term corresponding to hopping between nearest neighbor sites
and the second term to the onsite Coulomb repulsion provide the
competition between itineracy and localization in the regular Hubbard
model. The third term represents Coulomb repulsion between electrons
occupying nearest neighbor sites. The Hamiltonian as written above
guarantees an insulating ground state with a filling of one electron
per site. 
 
Although this model has been widely studied, many questions remain
unanswered. The analytic approach at its best in Bethe-Ansatz
provides an exact solution only for the $V=0$ case\cite{liebwu}. The
method has been used to obtain the energy spectrum and thermodynamics,
but a reliable computation of dynamical quantities such as the optical
response remains elusive. Non-perturbative analytic studies of the
dynamical response in these systems have largely been constrained to
use the continuum limit \cite{controzzi}.

Numerical methods such as exact diagonalization, although valuable in
providing real frequency information, are limited to small system
sizes \cite{fye}. Quantum Monte Carlo methods can treat large finite
size clusters but analytic continuation from imaginary to real
frequencies is an unreliable procedure. 
 
The renormalization group idea has helped deal with some of the
toughest problems in physics characterized by large number of degrees
of freedom playing an essential role. The efficacy of the idea as a
tool to compute experimentally relevant quantities lies in the ability
to integrate out the non essential degrees of freedom. This has been
brought to fruition with tremendous success in a numerical algorithm
for low dimensional interacting systems known as the Density Matrix
Renormalization Group (DMRG) \cite{white}. The numerical
solution of finite size systems on a computer is restricted by an
exponentially increasing Hilbert space. The DMRG method works around
this by a systematic truncation. It prescribes how to retain the most
probable states of a subsystem required for an accurate description of
a particular set of states (usually, just the ground state) of the
full system. The DMRG algorithm initially suited to deal only with
ground state properties, was subsequently extended to compute
dynamical correlation functions \cite{hallberg,kuhwhite}.  One of the
approaches known as the ``Lanczos vector method'' constitutes choosing
the particular set above as the ground state plus a set of Lanczos
vectors. The vectors are chosen to approximate the reduced Hilbert
space of excited states which connect to the ground state via the
operator whose correlation function is desired. This method is very
efficient in capturing low energy sharp features such as excitons in
the optical spectrum; especially when bulk of the weight is in a single
peak. Excitonic features in multi particle correlation functions
have been observed in the EHM in previous studies \cite{stephan,gallinar}.

The EHM at half-filling shows an interesting phase diagram. In the
weak coupling limit ($U \ll t$) the system undergoes a second order
transtion from a spin density wave (SDW) phase to a charge density
wave (CDW) state as a function of increasing $V$, at
$V=U/2+\delta(U)$. For intermediate values of $U$ the SDW and CDW
phases are separated by a narrow region with a bond charge density
wave (BCDW) order. As one approaches strong coupling ($U \gg t$), the
transition is again from an SDW to a CDW phase at $V=U/2+\delta(U)$,
but it is now first order. The small correction $\delta(U)$ is
positive and approaches zero at both the weak and strong coupling
ends. The precise location of a tricritical point at the crossover
between the first and second order transitions has been a subject of
much investigation \cite{hirsch,cannon,voit,zhang,nakamura} and is
complicated by the existence of the BCDW order.

In our study we compute the optical response and local spectral
function of the EHM in the strong coupling regime. We fix $U$ at a
realistic value of $12t$. The first order transition between the SDW
and CDW phases is manifest in the optical properties as well as in the
ground state energy. The relevant values of $V$ for SrCuO$_2$ and Ni
halides are in the SDW phase. All the results presented in this letter
were obtained from computations performed with finite size chains of
$N_s=50$ sites with open boundary conditions using the Lanczos vector
method. Studying other sizes ($N_s=18,34,66$) shows that results
for $N_s=50$ are generic. We use the finite size version of the DMRG
algorithm and choose $m=150$ for the largest sizes. Selected runs
performed with higher values of $m$ did not introduce significant
changes in the results. Typical discarded weights were
$O(10^{-6})$. To validate our code we compare our results for the
static and dynamic properties with exact diagonalization for short
chains.

We use the following definition for the response function,
\begin{equation}
\chi_{AB}(\omega)= \frac{i}{L} \int_{0}^{\infty} dt e^{i(\omega+i\epsilon)t}
\langle 0 \left| \left[ A^{\dagger}(t),B(0) \right] \right| 0 \rangle
\end{equation}
The real part of the optical conductivity is defined through the imaginary 
part of the current--current response with $\epsilon \rightarrow 0$,
\begin{equation}
\sigma'(\omega )=\frac{1}{\omega } \chi''_{jj}(\omega) 
\end{equation}
where  
\begin{equation}
j = -it\sum_{j,\sigma}\left( c_{j+1\sigma}^{\dagger }c_{j\sigma}
                            -c_{j\sigma}^{\dagger }c_{j+1\sigma}\right)
\end{equation}
is the paramagnetic current operator. The non-resonant Raman spectrum
in a Mott-Hubbard system is given by the response function of the
stress energy tensor \cite{shastry}, 
\begin{equation}
\tau  = -t\sum_{j,\sigma}\left( c_{j+1\sigma}^{\dagger }c_{j\sigma}
                               +c_{j\sigma}^{\dagger }c_{j+1\sigma}\right)
\end{equation}

In the case of an insulator $\chi_{\tau\tau}$ is not the dominant
contribution to the total Raman spectrum. But it is interesting for a
comparison with $\chi_{jj}$ because, $j$ and $\tau$ are respectively
odd and even under parity conjugation, apart from an overall phase.

\begin{figure}[t]
\centering \epsfig{file=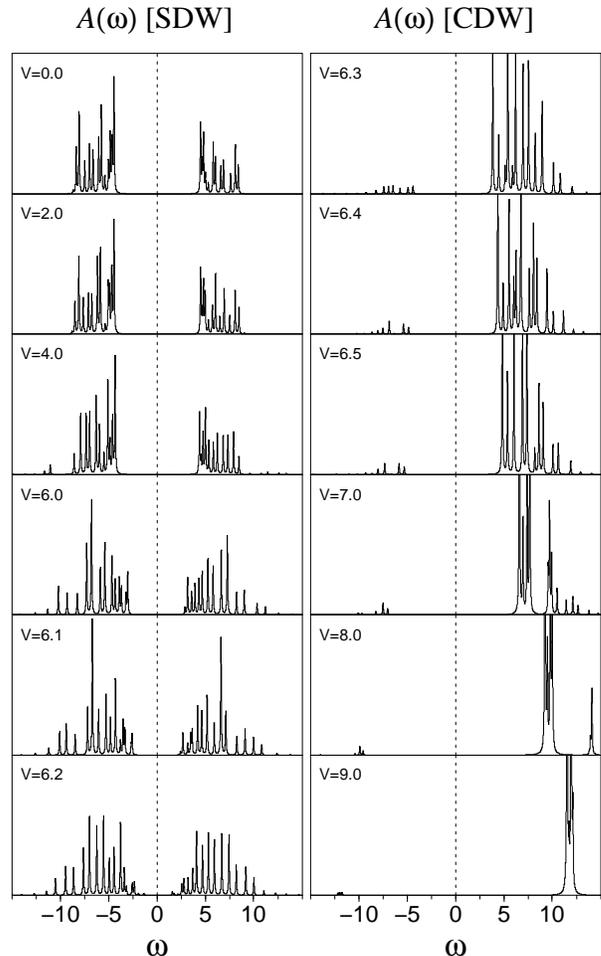}
\caption{Local Spectral functions for different values of the coupling $V$.
Note the abrupt change in particle--hole symmetry between the two phases.}
\label{spectral24}
\end{figure}

\begin{figure}[t]
\centering \epsfig{file=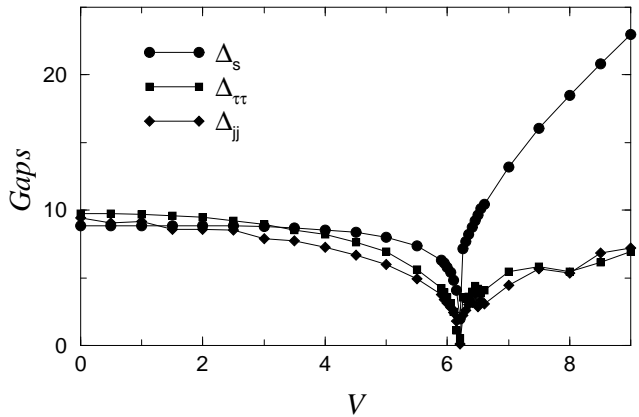,width=3.3in}
\caption{Gaps for different values of the coupling $V$.}
\label{gaps24}
\end{figure}

In Fig.\ref{spectral24} we show the local spectral function for
various values of $V$ ranging from $0$ to $9t$. In the SDW phase, the
single particle gap ($\Delta_s$) stays constant until a threshold
value of $V$ around $3t$ is reached and then starts reducing
(cf. Fig.\ref{gaps24}). Here and further in this letter when we refer
to the gap in a correlation function we measure the position of the
lowest energy peak. We ignore the tail part which comes about due to
the small finite broadening that is used to represent the Lanczos
continued fraction. This implies that our values for the gaps are
tight upper bounds to the actual ones. At the SDW-CDW transition the
spectral gap reduces abruptly reaching a finite value, jumps up and then
starts increasing again in the CDW phase. Note that in the CDW phase,
the site on which we compute the spectral function is empty in the
ground state.

\begin{figure}[t]
\centering \epsfig{file=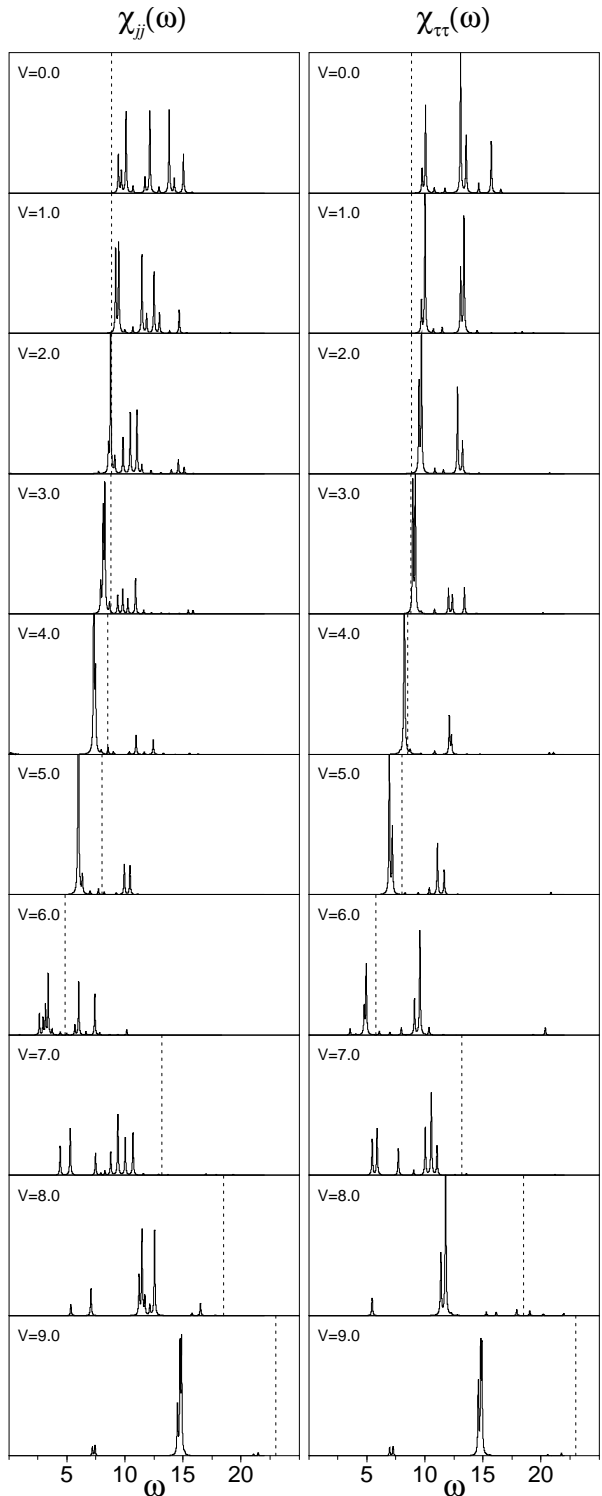}
\caption{Optical and Raman response functions for different values of
the coupling $V$. The vertical dotted line indicates the magnitude of
the single--particle spectral gap in each case.}
\label{resp24}
\end{figure}

In Fig.\ref{resp24} we report the current-current and non-resonant
Raman response functions in the left and right columns respectively. A
systematic change in the optical response in the odd ($\chi_{jj}$) and
even ($\chi_{\tau\tau}$) channels is discernable as we sweep through
$V$ across the SDW and into the CDW phase. For $V=0$ we see a broad
feature centered at around $U$, in good agreement with a recent
calculation for the standard Hubbard model within the DMRG
approach\cite{jeckelmann}. Our method does not allow a resolution of
the tiny bump seen in the middle of the broad optical absorption band.
But, as $V$ is increased we do notice the formation of a resonance
which gradually gains in weight and shifts towards lower
frequencies\cite{gallinar,gebhard}. This constitutes a precursor of
the excitonic feature that we describe further below.

For small values of $V$ the optical ($\Delta_{jj}$) and Raman
($\Delta_{\tau\tau}$) gaps would be expected to coincide with the
spectral gap, $\Delta_s$. We find them to be slightly larger because
it is not possible to create fully non-interacting electron--hole
pairs in a finite size chain. Beyond $V\sim1.5t$, the optical gap
falls below the spectral gap. The same happens for the Raman gap
around $V\sim3t$. We define a quantity that we will call the excitonic
weight ($W_{jj(\tau\tau)}$) as the fraction of the weight in the
optical (Raman) spectrum below $\Delta_s$;

\begin{equation}
W_{jj(\tau\tau)} = \frac{\int_{0}^{\Delta_s} d \omega~\chi_{jj(\tau\tau)}(\omega)}
              {\int_{0}^{\infty} d \omega~\chi_{jj(\tau\tau)}(\omega)}
\end{equation}

\begin{figure}[t]
\centering \epsfig{file=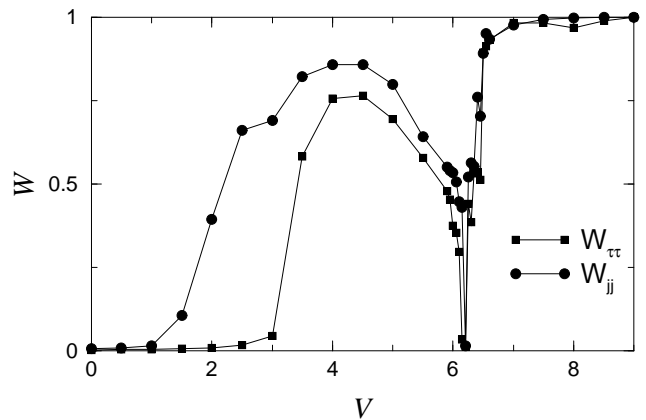,width=3.3in}
\caption{Excitonic weight as a function of the coupling $V$.}
\label{weight24}
\end{figure}

In Fig.\ref{weight24} we plot the excitonic weight as a function of
$V$. In the case of both response functions it is seen that the
excitonic weight is zero until the above mentioned crossing of gaps
occurs (cf. Fig.\ref{gaps24}). As $V$ is increased further, the
excitonic weight starts appearing and a resonance begins separating
from the rest of the spectrum. When $W_{jj(\tau\tau)}$ reaches a
maximum around $V\sim4t$, the spectrum is dominated by a sharp
excitonic feature carrying most of the weight --$86\%$($77\%$). This
peak, well differentiated from the rest of the spectrum, is clearly
located inside the single--particle spectral gap while the rest of the
weight falls outside. The Lanczos method is rather well suited to
describe these excitonic features, but it is not so good in capturing
detail at the higher end of the spectrum. The optical bands inside the
one--particle continuum tend to be shifted towards higher energies.
At the same time the relative weight in the excitonic features is
accurately represented, since we find that the sum rule for the
optical conductivity in terms of kinetic energy is obeyed with $1\%$
accuracy or better (except very close to the transition). As $V$ is
increased further beyond $V\sim4t$ the excitonic feature starts
loosing weight and at the same time marches towards zero frequency. At
the precise point of the SDW-CDW transition the excitonic mode reaches
the lowest frequencies we can resolve ($\omega\sim 1/L$).

In a Mott insulator represented by the half--filled Hubbard model,
creation of an independent electron--hole pair (or a holon--antiholon
pair in the Bethe Ansatz language) has a finite energy threshold;
namely the spectral gap. This threshold is lowered in the presence of
an attractive force by the binding energy of an electron--hole pair
called an exciton. This attraction comes about due to the increased
range of Coulomb repulsion in the EHM and is significantly absent in
the standard Hubbard model. As we increase $V$ the energy gained in
binding electron--hole pairs keeps growing continuously and the
excitonic feature moves closer and closer to zero frequency. At the
same time the density of electron and hole states available for
binding first increases, reaches a maximum and then goes to zero at
the SDW-CDW boundary. When the energy gained from binding the
electron--hole pairs equals the energy cost of creating them across
the single particle gap and the optical gap vanishes in both the odd
($\chi_{jj}$) and even ($\chi_{\tau\tau}$) channels.

Although our interest is in the strong coupling limit one can gain an
understanding of the physics at hand by using the language of the
continuum limit applied to the EHM\cite{voit}; an approach which is
strictly valid only in the weak coupling case. Further using
bosonization, the model can be split up into two Sine Gordon Models
(SGM) for the charge and spin sectors respectively. The charge sector
is the only one of interest as far as ($\chi_{jj}$) is concerned. On
the other hand spin--charge separation does not help a study of
($\chi_{\tau\tau}$) because the operator $\tau$ involves both the
charge and spin sectors. The exact solution of the SGM is available
and the spectrum is governed by a coupling parameter usually called
$\beta$\cite{gogolin}. For $\beta^2<4\pi$ the spectrum is built solely
of kinks and antikinks. As $\beta$ increases further, the SGM enters
the attractive regime and kink--antikink bound states known as
breathers are formed. These bound states in the SGM correspond to the
excitons in the EHM that form as the value of $V$ is increased. Given
that we are interested in the strong coupling regime of the EHM, a
quantitative comparison with the attractive regime of the SGM falls
beyond the margin of applicability of the scaling limit.

To conclude, we have shown that sharp excitonic features dominate the
transport behavior in a particular regime of the Extended Hubbard
Model at half-filling. These excitons are a direct consequence of the
presence of non--local Coulomb interaction in this model. The Lanczos
method combined with the DMRG approach is a powerful non-perturbative
tool in computing the dynamical correlation functions of a non-trivial
system.  Several materials such as SrCuO$_2$, Ni halides and the
(TMTTF)$_2$ X salts (where X is an inorganic monoanion) are believed
to be good examples of Mott insulators\cite{mizuno,kishida,tmttf}. Our
numerical approach permits us to easily include other ingredients such
as explicit dimerization and interchain hopping which are present in
these materials in order to allow a quantitative comparison with
experiments in the future.

We are indebted to G. Kotliar for several suggestions and to A. Rosch
for his keen interest and many comments.  We acknowledge useful
discussions with N. Andrei, A. Millis and S. Shastry. We thank
K. Hallberg and S.R. White for discussions on the DMRG method.

\end{multicols}

\end{document}